\documentclass[prb,reprint,superscriptaddress]{revtex4-1}

\usepackage[T1]{fontenc}
\usepackage[utf8]{inputenc}
\usepackage{graphicx}
\usepackage{amsmath}
\usepackage{color}
\usepackage{comment}
\usepackage{hyperref}
\usepackage{multirow}
\usepackage{subfig}
\usepackage{braket}
\usepackage{bm}
\usepackage[english]{babel}

\begin{document}

\title{Ground state of Fe(II)-porphyrin model  corresponds to quintet: A DFT and DMRG-based tailored CC study}

\author{Andrej Antal\'{i}k}
\affiliation{J. Heyrovsk\'{y} Institute of Physical Chemistry, Academy of Sciences of the Czech \mbox{Republic, v.v.i.},
Dolej\v{s}kova 3, 18223 Prague 8, Czech Republic}
\affiliation{Faculty of Mathematics and Physics, Charles University, Ke Karlovu 3, 12116, Prague 2, Czech Republic}

\author{Dana Nachtigallov\'{a}}
\affiliation{Institute of Organic Chemistry and Biochemistry, Academy of Sciences of the Czech \mbox{Republic, v.v.i.}, Flemingovo n\'{a}m. 2, 16610 Prague 6, Czech Republic}
\affiliation{Regional Centre of Advanced Technologies and Materials, Palack\'{y} University, 77146 Olomouc, Czech Republic}

\author{Rabindranath Lo}
\affiliation{Institute of Organic Chemistry and Biochemistry, Academy of Sciences of the Czech \mbox{Republic, v.v.i.}, Flemingovo n\'{a}m. 2, 16610 Prague 6, Czech Republic}
\affiliation{Regional Centre of Advanced Technologies and Materials, Palack\'{y} University, 77146 Olomouc, Czech Republic}

\author{Mikul\'{a}\v{s} Matou\v{s}ek}
\affiliation{J. Heyrovsk\'{y} Institute of Physical Chemistry, Academy of Sciences of the Czech \mbox{Republic, v.v.i.},
Dolej\v{s}kova 3, 18223 Prague 8, Czech Republic}
\affiliation{Faculty of Mathematics and Physics, Charles University, Ke Karlovu 3, 12116, Prague 2, Czech Republic}

\author{Jakub Lang}
\affiliation{J. Heyrovsk\'{y} Institute of Physical Chemistry, Academy of Sciences of the Czech \mbox{Republic, v.v.i.},
Dolej\v{s}kova 3, 18223 Prague 8, Czech Republic}

\author{\"Ors Legeza}
\affiliation{Strongly Correlated Systems ``Lend\"{u}let'' Research group, Wigner Research Centre for Physics, H-1525, Budapest, Hungary}

\author{Ji\v{r}\'{i} Pittner}
\affiliation{J. Heyrovsk\'{y} Institute of Physical Chemistry, Academy of Sciences of the Czech \mbox{Republic, v.v.i.}, Dolej\v{s}kova 3, 18223 Prague 8, Czech Republic}

\author{Pavel Hobza}
\email{pavel.hobza@uochb.cas.cz}
\affiliation{Institute of Organic Chemistry and Biochemistry, Academy of Sciences of the Czech \mbox{Republic, v.v.i.}, Flemingovo n\'{a}m. 2, 16610 Prague 6, Czech Republic}
\affiliation{Regional Centre of Advanced Technologies and Materials, Palack\'{y} University, 77146 Olomouc, Czech Republic}

\author{Libor Veis}
\email{libor.veis@jh-inst.cas.cz}
\affiliation{J. Heyrovsk\'{y} Institute of Physical Chemistry, Academy of Sciences of the Czech \mbox{Republic, v.v.i.}, Dolej\v{s}kova 3, 18223 Prague 8, Czech Republic}

\newcommand{\ph}[1]{\phantom{#1}}
\newcommand{\EQT}{$\Delta E^\mathrm{Q \rightarrow T}$}

\maketitle

\textbf{Fe(II)-porphyrins play an important role in many reactions relevant to material science and biological
  processes, due to their closely lying spin states.
  %However, this small energetic separation also makes it challenging to establish the correct spin state ordering.
  Although the prevalent opinion is that these systems posses the triplet ground state, the recent experiment
  on Fe(II)-phthalocyanine under conditions matching those of an isolated molecule points toward the quintet
  ground state.
  We present a thorough DFT and DMRG-based tailored CC study of Fe(II)-porphyrin model, %by means of the density functional theory and density matrix
  %renormalization group based tailored coupled clusters, 
  in which we address all previously discussed
  correlation effects.
  We examine the importance of geometrical parameters, the Fe--N distances in particular, and conclude that
  the system possesses the quintet ground state.} \\ %, which is in our calculations well-separated from the triplet state.

\begin{figure}[!ht]
	\centering
	\includegraphics[width=0.5\textwidth]{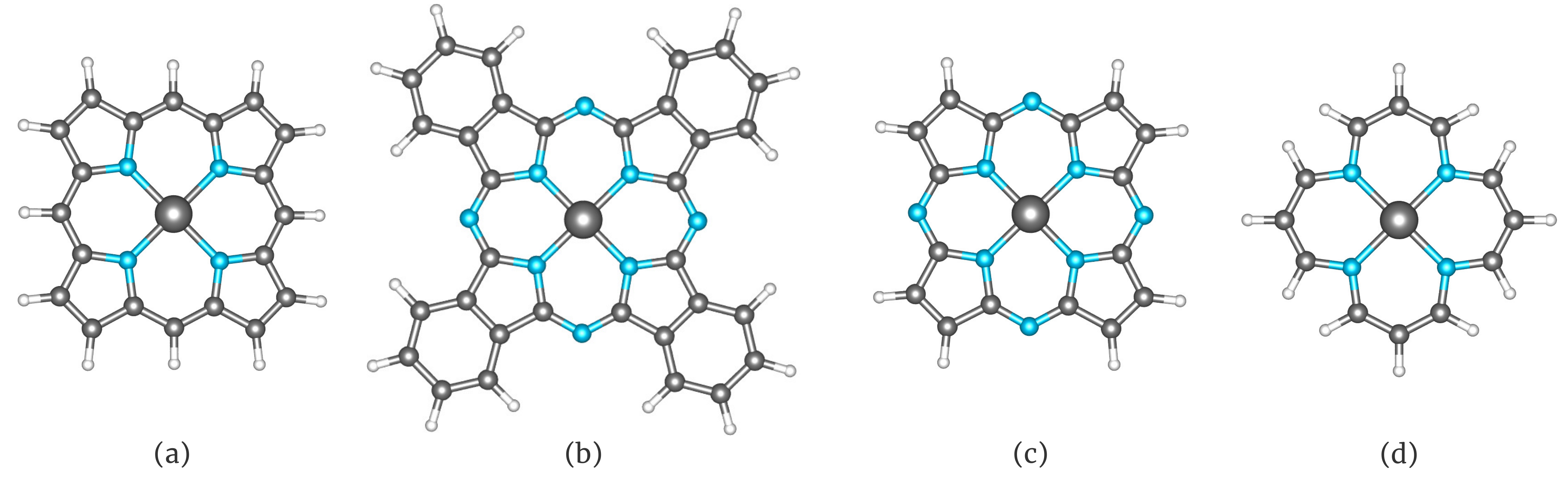}
  \caption{Structures of Fe(II)-porphyrin (a), Fe(II)-phtalocyanine (b), Fe(II)-porphyrazine (c), and a model system of Fe(II)-porphyrin (d) studied in this work and in the text denoted as {\bf 1}, or depending on the geometry, the Fe-N distance in particular, employed {\bf 1a}, {\bf 1b}, and {\bf 1c} (see Table \ref{tab:geom}).}
  %Color key: iron (gray, large), nitrogen (blue), carbon (gray, small), hydrogen (white).}
  \label{fig:molecules}	
\end{figure}

%%% MAIN TEXT %%%
%%% INTRO %%%%
Porphyrins are conjugated aromatic systems composed of four pyrrole rings connected at their C$_\alpha$
atoms by C$_\beta$H groups (see Fig. \ref{fig:molecules}).
Their metal-derivates, in particular Fe(II)-porphyrins based on Fe(II)-porphyrin (FeP, Fig. \ref{fig:molecules}a)
(Fe(II)-phtalocyanine (FePc, Fig. \ref{fig:molecules}b) and Fe(II)-porphyrazine (FePz, Fig. \ref{fig:molecules}c)),
play an important role in reactions related to material science and biological processes due to the near degeneracy
of their high-spin (quintet), intermediate-spin (triplet) and low-spin (singlet) states.
A well-known example is the triplet to singlet spin crossover upon binding of molecular oxygen to the Fe(II)
active site of hemoglobin\cite{Kepp2017}.
%Despite the huge efforts over the years, the exact mechanism of this reaction has not yet been established.

%The major obstacle in the investigation of the heme molecule (especially hemoglobin) is its size, which
%renders the use of high-level calculations, necessary for definitive assignment of its ground state, challenging.
%For this reason, only the core formed by FeP is used to model this system.
%However, since it does not exist in an unsubstituted state, similarly to FePz, the reliability of computational
%methods based on the comparison with experiment cannot be assessed.
%Existing experimental studies of four-coordinated Fe(II) embedded in substituted porphyrin systems
%\cite{Spartalian1979,Dale1968,Evangelisti2002,Filoti2006,Bartolom2010,Gruyters2012,
%Kitagawa1979,Collman1975,Goff1977,Mispelter1980}
%have assigned mostly the triplet ground state, with the exception of the high-spin quintet ground state of
%Fe(II)-octametyl-tetrabenzporphyrin that has been assigned by the M\"{o}ssbauer spectroscopy\cite{Sams1974}.
%These experiments have been performed either in the crystal phase or in polar solvent, which might affect
%the ground state and currently make highly correlated calculations not directly comparable with experiment.
%Gas phase studies in which an isolated molecule is investigated were limited to high temperatures which alter
%the ordering of closely lying spin states.

Since neither FeP, nor FePz exists in an unsubstituted state, the direct comparison of theoretical and experimental results
is not possible.
Although the existing experimental studies on four-coordinated Fe(II) embedded in substituted porphyrin systems
\cite{Spartalian1979,Evangelisti2002,Filoti2006,Bartolom2010,Gruyters2012, Kitagawa1979,Collman1975,Mispelter1980, Sams1974}
mostly predict the triplet ground state, they have been performed either in the crystal phase or polar solvent, 
which are far from the gas phase conditions of computational studies.

We have recently investigated the effect of different environments on the FePc ground state
%The effect of different environments on the FePc ground state was recently investigated in our laboratory
by means of the M\"{o}ssbauer spectroscopy and various computational methods \cite{Nachtigallov2018}.
%a combined experimental and computational study.\cite{Nachtigallov2018}.
%We measured M\"{o}ssbauer spectra in crystalline form and in the dimethyl-formamide (DMF) and
%monochlorobenzene solvents, which are characterized by different dielectric constants (D$_e$) and dipole moments
%($\mu_\text{dip}$), (DMF: D$_e \sim 37$, $\mu_\text{dip}$= 3.9  D;
%monochlorobenzene: D$_e \sim 6$, $\mu_\text{dip}$ = 1.5 D).
%The low solubility and polarity of monochlorobenzene closely resemble the gas phase conditions of 
%computational studies; the isolated nature of FePc is confirmed also by the zero Weiss temperature.
%The observed characteristics, i.e. the isomer shift ($\delta$) and quadrupole splitting ($\Delta E_\text{Q}$) 
Our experiments 
have
unambiguously indicated the triplet ground state of FePc in the crystalline form and dissolved in dimethyl-formamide,
and the quintet when dissolved in monochlorobenzene (resemblance to gas phase conditions of 
computational studies).
The quintet ground state was also confirmed by the multireference density matrix renormalization group (DMRG)
calculations.
These findings clearly contradict the prevalent opinion in the literature.

The experimental observations of Fe(II)-porphyrins guided several follow-up computational studies
on FeP and FePz with the density functional theory (DFT)
\cite{Bartolom2010,Nachtigallov2018,Kroll2012,Kuzmin2013,Berryman2015,Liao2006,
Groenhof2005,Walker2010,Brena2011,Marom2008,Obara1982,Kuzmin2009,Reynolds1991,Liao2001} and multireference
methods\cite{Ichibha2017,Pierloot2017,Phung2018,Sun2017,LiManni2016,Smith2017,LiManni2018,LiManni2019,Radon2008}.
%Depending on the functional, DFT calculations have assigned either quintet or more frequently triplet
%ground-state for FeP\cite{Berryman2015,Liao2006,Groenhof2005,Walker2010,Liao2001}.
%On the contrary, the complete active space perturbation theory (CASPT2)\cite{Radon2008} has predicted
%the quintet ground-state which has lead to the discussion of various computational aspects, mainly the
%Fe(3s,3p) electron correlation \cite{Pierloot2017}.
%Its evaluation with CASPT2 and coupled clusters (CC) with singles, doubles and perturbative triples (CCSD(T))
%puts triplet 0.15 kcal/mol below quintet\cite{Phung2018}.
Nevertheless, due to the complexity of the problem, an unambiguous answer to the state ordering has not been found even with
the multireference methods that are able to correlate a relatively large number of electrons \cite{OlivaresAmaya2015,Zhou2019,LiManni2016,Smith2017}.
%DMRG \cite{Sun2017,OlivaresAmaya2015,Zhou2019},
%full configuration interaction quantum Monte Carlo\cite{LiManni2016}
%or Heat-bath Configuration Interaction algorithm coupled with the multiconfiguration self-consistent
%field\cite{Smith2017} methods, which are able to correlate a relatively large number of electrons.

%It should be noted that due to the size of FePc neither the DMRG calculations provide a full description
%of the electron correlation\cite{Nachtigallov2018}, although they predict the quintet ground state
%in agreement with the experiment.

The effects of different contributions to electron correlation,\cite{Pierloot2017,LiManni2018,Phung2018}
%in particular previously discussed semi-core Fe(3s,3p) electron correlation between the Fe and
%the $\pi$-space of the porphyrin ring\cite{Pierloot2017} and out-of-plane Fe(3d) orbitals\cite{LiManni2018},
%as well as electron correlation beyond the active space\cite{Phung2018}
have recently been investigated by calculations on model system of Fe(II)-porphyrin in which the
bridging C$_\beta$H groups are replaced by hydrogen atoms (model {\bf 1}, Fig. \ref{fig:molecules}d).
In the recent work of Li Manni et al.\cite{LiManni2019}, the complete active space (CAS) was constructed
from 32 electrons in 34 orbitals, in particular, the Fe(3d), Fe(4d), $\sigma$ lone pairs, and all
$\pi$ orbitals of the porphyrin model ring to cover the valence correlation.
%These orbitals describe the $\sigma$-donation/$\pi$-back-donation interaction with the ring, which significantly
%contributes to the triplet stabilization with respect to quintet.
The active space was then augmented by the semi-core Fe(3s,3p) orbitals resulting in CAS(40,38)
and a minor increase in the quintet-triplet gap.
The inclusion of beyond-CAS correlation by employing the single reference coupled cluster correction
further stabilized the triplet ground state and provided the final estimate of the triplet-quintet energy gap
as 5.7 kcal/mol.
Comparing these results to the FePc experimental data\cite{Nachtigallov2018} and our preliminary DFT calculations
on the spin state ordering in FeP and FePc (see ESI), we reopen the debate over the character of the
Fe(II)-porphyrins and their modelling.
Besides discussing the extent of electron correlation in the multireference approach, we explore
the role of other parameters which may influence the ground state predictions.

Among such parameters, particular attention should be paid to the geometry of FeP systems.
The Fe--N bond distance ($R_\text{FeN}$) has been discussed by several
authors\cite{Choe1999,Sontum1983,Nagashima1986}, with some proposing that the increase in $R_\text{FeN}$
stabilizes quintet states via the relaxation of d$_{\text{x}^2-\text{y}^2}$ orbital\cite{Choe1998}.
The calculated Fe--N bond distances obtained for the quintet states typically range from 2.0 to 2.1 {\AA}
\cite{Choe1999,Rovira1997,Vancoillie2011}.
In comparison, the value of 1.972 {\AA} taken from the X-ray diffraction of Fe(II)-tetraphenylporphyrin
(FeTPP)\cite{Collman1975} is closer to the value of 1.989 {\AA} obtained for the FeP triplet state by employing
the PBE0 functional in DFT optimization\cite{Vancoillie2011}.
This result confirms the suggestion discussed in Ref.\cite{Nachtigallov2018} for Fe(II)-Pc, according to which
the ground spin state observed in the crystalline form of Fe(II)-porphyrins very likely differs from the ground
state of an isolated molecule in the gas phase.

This discussion on various effects influencing the spin state ordering raises the following question:
Does the improved electron correlation treatment result in the same changes in the triplet-quintet state ordering
of the FeP model regardless of whether the triplet optimized or quintet optimized distance is used?
%of the FeP model regardless of whether the triplet optimized ($R_\text{FeN}=1.989$ {\AA}\cite{LiManni2019}) or
%quintet optimized ($R_\text{FeN}>2$ {\AA}) distance is used?

%%% METHODS %%%
To resolve this issue, apart from the DFT calculations with the B97-D3 functional, we investigated the
electronic structure of {\bf 1} by means of DMRG-based methods.
%To resolve this issue, apart from the DFT calculations, we investigated the electronic structure of {\bf 1} by means
%In this communication, apart from the DFT calculations, we investigated the electronic structure of {\bf 1} by means
%of the state-of-the-art multireference methods, namely DMRG and its extension in the form of tailored coupled clusters (TCC) \cite{Veis2016}.
%Here, we only outline the main ideas behind the DMRG-TCC method as its detailed description and mathematical
%background is available in one of our previous works\cite{Veis2016,Faulstich2019,Faulstich2019_n2}.
DMRG is a well-established and very powerful approach suitable for treatment of strongly correlated problems that
require large active spaces \cite{White1992,Chan2011}.
However, despite its favorable scaling, it is still computationally prohibitive to treat the dynamic correlation
solely with DMRG. %i.e. to include all virtual orbitals into the active space.
As a possible solution, we have introduced the TCCSD method, in which the CC wave function is externally
corrected using the information extracted from the DMRG calculation\cite{Veis2016}.
We showed that it is able to describe both non-dynamic and dynamic correlation in a balanced way\cite{Veis2018},
but due to the scaling of the CCSD part, the TCCSD methodology quickly becomes unfeasible for larger systems.
To remove this bottleneck, we have recently developed its domain-based local pair natural orbital
(DLPNO) version\cite{Lang2020}, which employs the pair natural orbitals to exploit the locality of electron
correlation\cite{Neese2009b,Neese2009a,Riplinger2013}.
The electronic structure of parent FeP was recently studied also by means of an alternative post-DMRG method, namely the DMRG-based pair density functional theory (DMRG-PDFT) \cite{Zhou2019}.

%These are natural orbitals specific for each pair of localized occupied orbitals
%and which are known to provide compact parametrization of the virtual orbital space\cite{Neese2009b,Neese2009a}.
%Compared to the other nearly linear scaling methods, the main advantage of the DLPNO formalism\cite{Riplinger2013}
%is that it is controlled only by a limited number of cut-off parameters which do not explicitly involve
%distances in real space.

% DORIESIT SKOK + ODSTAVEC

% DFT
The $R_\text{FeN}$ values resulting from spin separate triplet and quintet optimizations of model {\bf 1},
performed at the B97-D3/def2-TZVPP level, are given in Table \ref{tab:geom}.
For comparison, we also report the distances for FeP and FeTPP, which are in very good agreement with the PBE0
values of 1.989 {\AA} and 2.053 {\AA} optimized for the FeP triplet and quintet states,
respectively\cite{Vancoillie2011}.
Additionally, the $R_\text{FeN}$ values obtained from the triplet optimizations agree reasonably well with
the distance of 1.972 {\AA} found in the X-ray diffraction experiment\cite{Collman1975} (where FeTPP is predicted
to possess the triplet state), thus confirming the reliability of B97-D3 functional.
In agreement with the discussion above, the quintet-optimized $R_\text{FeN}$ values of FeP, FeTPP and {\bf 1}
are larger compared to the triplet state, with the differences 0.060 {\AA}, 0.067 {\AA} and 0.132 {\AA},
respectively.
The significant increase in elongation for {\bf 1} compared to FeP and FeTPP stems from the larger flexibility
of the surrounding ring because of the missing bridging C$_\beta$H groups.

\begin{table}
  \small
  \caption{The Fe--N distance ($R_\text{FeN}$, in \AA) optimized for each state at the B97-D3/def2-TZVPP level for
  FeP-based systems with the exception of {\bf 1a} which is from Ref.\cite{LiManni2019}}
  \label{tab:geom}
  \def\arraystretch{1.2}
  \begin{tabular*}{0.48\textwidth}{@{\extracolsep{\fill}}lll}
  \hline
                        & Triplet & Quintet \\ \hline
   FeP model (\bf{1})   &  1.989 (\bf{1a}) &      \\
                        &  2.048 (\bf{1b}) & 2.180 (\bf{1c})      \\ \hline
   FeP                  &  1.997 &  2.064                \\ \hline %(FeP(Q)) \\ \hline
   FeTPP                & 1.995, 1.998 &  2.063--2.065         \\ \hline
  \end{tabular*}
\end{table}

Table \ref{tab:dft} lists the relative spin state energies from DFT obtained by employing the B97-D3 functional
for various geometries of {\bf 1} and their comparison with the previously reported results on {\bf 1a}
obtained with the Stochastic-CASSCF\cite{LiManni2018,LiManni2019}.
The B97-D3 adiabatic energy gap is determined as 11.0 kcal/mol with the quintet ground state by using
the Fe--N distances from the fully optimized triplet {\bf 1b} and quintet {\bf 1c} geometries
(the difference in $R_\text{FeN}$ is 0.132 {\AA}).
This gap then reduces to 2.8 kcal/mol when $R_\text{FeN}$ values from the optimized FeP are used
(the difference in $R_\text{FeN}$ is 0.067 {\AA}).
The vertical gap at the triplet geometry {\bf 1b} results in reversed ordering with the triplet state more stable
than quintet by 2.9 kcal/mol and it increases to 8.0 kcal/mol when $R_\text{FeN}$ optimized for FeP triplet is used.
At the similar Fe--N distance {\bf 1a}, the Stochastic-CASSCF calculations\cite{LiManni2018,LiManni2019} predict
the triplet ground state as well, but with the smaller energy gap of 3.1 and 4.4 kcal/mol
using the CAS(32,34) and CAS(40,38), respectively. 

\begin{table*}
  \small
  \caption{Relative energies in kcal/mol of the triplet and quintet states of {\bf 1} based on the DFT calculations and
  Stochastic-CASSCF calculations from Ref.\cite{LiManni2019}.
  Geometry denotes the source of geometry parameters}
  \label{tab:dft}
  \def\arraystretch{1.2}
  \begin{tabular}{lcccc}
    \hline
    Method                   & Excitation & Geometry           & Triplet & Quintet \\ \hline
    B97-D3/def2-TZVPP        & adiab.     & {\bf 1b}, {\bf 1c} & 11.0    & 0.0 \\
                             & vert.      & {\bf 1b}           & 0.0     & 2.9 \\
                             & adiab.     & FeP (T,Q)          & 2.8     & 0.0 \\
                             & vert.      & FeP (T)            & 0.0     & 8.0 \\ \hline
     Stoch.-CASSCF(32,34)    & vert.      & {\bf 1a}           & 0.0     & 3.1 \\
    \ph{Stoch.-CASSCF}(40,38)& vert.      & {\bf 1a}           & 0.0     & \ph{$^a$}4.4$^a$ \\ \hline
    \multicolumn{5}{l}{$^a$\,{\footnotesize Involving CCSD(T) correlation treatment increases the gap to 5.7 kcal/mol.}}
  \end{tabular} 
\end{table*}

% TOTO
These results indicate that the Fe--N bond distances play a significant role in the spin state ordering
of FeP systems, but the extent of its influence has not yet been evaluated in detail.
In fact, it seems that the value of this structure parameter can dominate the energy balance
and thus relative ordering of the spin states.
We evaluate this effect together with another significant influence which is the level of electron correlation
treatment.
In the following, we present the main results of (DMRG-)CASSCF and TCC calculations, while the complete
set of energies together with Computational Details is provided in the ESI.

% PREMOSTENIE na DMRG-TCC a ELECTRONIC STRUCTURE
Based on the previous discussions on the ground state of FeP systems in literature, only the lowest
quintet ($^5$A$_{1\text{g}}$) and triplet states are considered.
Unlike in the study of Li Manni et al.\cite{LiManni2019}, the lowest triplet state in all our (DMRG-)CASSCF
and TCC calculations corresponds to $^3$A$_{2\text{g}}$ with the occupation
(d$_{\text{x}^2-\text{y}^2}$)$^2$(d$_{\text{z}^2}$)$^2$(d$_\text{xz}$)$^1$(d$_\text{yz}$)$^1$(d$_\text{xy}$)$^0$.
Considering a very small energy gap of only about 0.5 kcal/mol between the two lowest triplets ($^3$A$_{2\text{g}}$
and $^3$E$_\text{g}$) in the aforementioned study, this discrepancy might be attributed to the difference
in basis sets.
Nevertheless, we believe that such a small energy gap is below the resolution of the employed methods.
Also, the $^3$A$_{2\text{g}}$ state was found to be the lowest triplet state of FeP in Ref. \cite{Vancoillie2011}.

% DLPNO accuracy
In order to assess the accuracy of the DLPNO approximation, we first performed a series of benchmark calculations.
In these, we calculated the energy differences of the studied quintet to triplet energy gaps
$\Delta E^\mathrm{Q\rightarrow T}=E^\text{T}-E^\text{Q}$
between the canonical TCC methods and its DLPNO counterparts in the smaller SVP basis set.
The resulting errors coming from the DLPNO approximation are well below 0.5 kcal/mol, except
those obtained for {\bf 1a} with CAS(8,12), where the errors are about 0.6 kcal/mol.

\begin{figure}
  \includegraphics[width=0.50\textwidth]{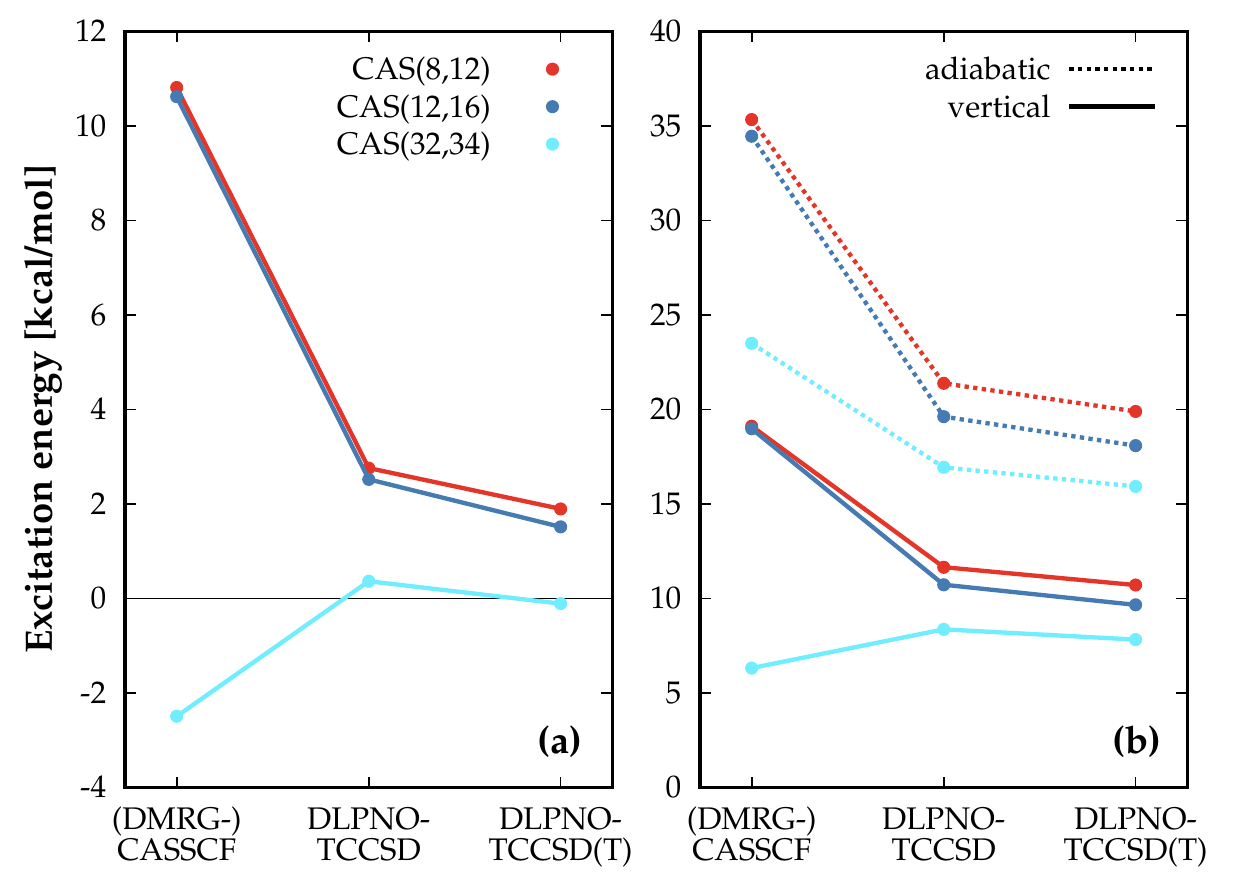}
  \caption{The (DMRG-)CASSCF, DLPNO-TCCSD, and DLPNO-TCCSD(T)
  (a) vertical \EQT energy gaps of {\bf 1a}\cite{LiManni2019},
  (b) vertical \EQT energy gaps of {\bf 1b} and adiabatic energy gaps of fully optimized {\bf 1}
  in kcal/mol in def2-TZVP basis. The keys are shared by both plots.}
  \label{fig:vertadiab}
\end{figure}

% 1a RESULTS - CAS(8,12) and CAS(12,16)
We first discuss the results for vertical \EQT in the {\bf 1a} geometry which are presented in Fig. \ref{fig:vertadiab}a.
This system has already been a subject of previous studies by Li Manni et
al.\cite{LiManni2018,LiManni2019} and it therefore offers an opportunity to compare our approach
with a different method.
Starting with the smaller CAS(8,12) and CAS(12,16), CASSCF results show an initial stabilization of the quintet state.
Similarly to Ref.\cite{LiManni2019}, the additional dynamic correlation on top of the CASSCF reference
wave functions by means of the DLPNO-TCCSD stabilizes the triplet, i.e. decreases \EQT.
Its further, yet less prominent stabilization is observed when perturbative triples are employed.
At this point, it is obvious that the inclusion of four Gouterman's $\pi$-orbitals\cite{Gouterman1961}
in CAS does not change the relative energies of the lowest quintet and triplet states and virtually
no difference in enegy gap between CAS(8,12) and CAS(12,16) at all levels of correlation treatment is observed.
However, the situation is different when the largest active space is used.

% 1a RESULTS - CAS(32,34)
While for the smaller active spaces each method assigns the quintet ground state, the addition of all
$\pi$-orbitals stabilizes the triplet state with respect to quintet at the DMRG-CASSCF(32,34) level.
Thus, the triplet becomes the ground state with \EQT gap corresponding to -2.49 kcal/mol, which agrees well with
the energy gap of -3.1 kcal/mol of the said study\cite{LiManni2018}.
The difference between these two values might originate in the use of different basis sets and/or slightly
differently optimized CASSCF orbitals, since the bond dimension in DMRG-CASSCF is not in full accordance with the given
number of walkers in Stochastic-CASSCF.
When the dynamic correlation is added on top of DMRG-CASSCF, the change in \EQT is less prominent compared to
the smaller active spaces.
This means that the majority of important correlations is already captured by the active space containing 34 orbitals
as previously discussed\cite{LiManni2018}.
Interestingly, the dynamic correlation stabilizes the quintet state, resulting in \EQT of $-0.11$ kcal/mol at our
highest level of theory DLPNO-TCCSD(T)(32,34), which contrasts with the previous observations
$-5.7$ kcal/mol\cite{LiManni2019}.
We would like to add that in Ref. \cite{Zhou2019}, the effect of active space on 
the energetic ordering of the lowest spin states of FeP was studied 
by means of DMRG-PDFT. The ground state was identified as triplet, the question of the geometry was not discussed in Ref. \cite{Zhou2019} though.

% DISCUSSION ad Li Manni
This inconsistency between our and the previously published results deserves a few comments.
In the study\cite{LiManni2019}, the authors investigated the effect of Fe(3s,3p) orbitals,
as well as the virtual orbitals not included in CAS(32,34).
Both groups of orbitals have a different stabilization effect: semi-core orbitals stabilize the triplet
state, whereas the dynamic correlation of the full virtual space stabilizes the quintet state.
In our opinion, the observed discrepancy stems from the fact that in the article by Li Manni et al.\cite{LiManni2019}
the correlation effects have been studied at a different level of theory.
The semi-core orbitals have been eventually included into the active space, and thus described at
the multireference level, while the effect of full virtual space has been studied by means
of single-reference CC.
Taking into account that in our TCC calculations the HF determinants contribute to the total wave function
with the weight of about 0.6, the single-reference level of theory might be inadequate.
Even though we employ the single-reference CC formalism (using one-determinant Fermi vacuum), our TCC approach
systematically accounts for the strong-correlation effects via the CC amplitudes extracted from the DMRG wave
function.
The semi-core correlation is included directly at the CCSD level and the effect of triplet stabilization is
even more prominent than in Li Manni et al.\cite{LiManni2019} (with respect to calculations with frozen
Fe(3s,3p) orbitals, see ESI).
Nevertheless, further studies which would employ alternative computational methods of calculation of dynamic
correlation on top of CASSCF(32,34) (e.g. adiabatic connection\cite{Pastorczak2018})
are necessary to confirm our hypothesis.

Since Rado\'{n} showed that CCSD(T) method itself can perform well on FeP \cite{radon}, we have carried out additional single reference DLPNO-CCSD(T) calculations of the FeP model. The resulting vertical gaps presented in ESI are in qualitative agreement with the DLPNO-TCCSD(T) results, differing by 2.8 and 3.6 kcal/mol in case of \textbf{1a} and \textbf{1b} geometry respectively.

\begin{comment}
\begin{figure}
	\centering
	\includegraphics[width=0.48\textwidth]{./new_1b1c.pdf}
  %\hfill
%	\includegraphics[width=0.48\textwidth]{./adiab.pdf}
  \caption{The (DMRG-)CASSCF, DLPNO-TCCSD, and DLPNO-TCCSD(T) vertical quintet to triplet energy gaps \EQT of
  {\bf 1b} (solid lines) and adiabatic energy gaps \EQT of fully optimized {\bf 1} (dashed lines) in kcal/mol in def2-TZVP basis.
  The triplet state optimized geometry {\bf 1b}: $R_\text{FeN}$ = 2.048 {\AA},
  the quintet state optimized geometry {\bf 1c}: $R_\text{FeN}$ = 2.180 {\AA}.}
  \label{fig:vertadiab}	
\end{figure}
\end{comment}

% 1b and ADIABATIC RESULTS
Next, the results are analyzed in terms of $R_\text{FeN}$ distance for spin state specific optimized
structures of FeP model {\bf 1} and presented in Fig. \ref{fig:vertadiab}b.
The solid lines show the values of vertical \EQT calculated for the {\bf 1b} geometry, which with improving treatment
of electronic correlation exhibit very similar trends as for {\bf 1a}, but shifted by about 8 kcal/mol towards
the more stable quintet.
The dashed lines show the values of adiabatic \EQT calculated for fully optimized {\bf 1} i.e. with the triplet
and quintet states in {\bf 1b} and {\bf 1c} geometries, respectively.
Compared to the vertical \EQT, these stabilize the quintet even more.

% BASIS SET
%The flexibility of the basis sets is demonstrated in Figure \ref{fig:vert_basis}, in which the results of vertical
%\EQT of {\bf 1a} and {\bf 1b} calculated with CAS(32,34) and def2-TZVP basis set are compared with those %obtained
%with def2-SVP basis set.
%The larger basis systematically stabilizes triplet with respect to the quintet state, the effect which
%is more prominent for {\bf 1a}.
%Even though the differences in \EQT values are small, ranging from 1.2 to 2.6 kcal/mol, they can change
%the ordering of spin states as observed for {\bf 1a}. 
%
%\begin{figure}
%	\centering
%	\includegraphics[width=0.48\textwidth]{./bases.pdf}
%  \caption{The (DMRG-)CASSCF, DLPNO-TCCSD, and DLPNO-TCCSD(T) vertical quintet to triplet energy gaps,
%  \EQT of {\bf 1a} ($R_\text{FeN}$ = 1.989 \AA) and {\bf 1b} ($R_\text{FeN}$ = 2.048 \AA) in kcal/mol using
%  def2-SVP and def2-TZVP basis sets.}
%  \label{fig:vert_basis}	
%\end{figure}

% DISCUSSION GEOMETRY
Now, considering the most important geometrical parameter $R_\text{FeN}$ of the models used in this study,
our best estimate of the vertical \EQT of {\bf 1a} ($R_\text{FeN} = 1.989$ {\AA}) obtained at the 
DLPNO-TCCSD(T)(32,34)/def2-TZVP level of theory puts the triplet state below quintet with the negligible gap
of $-0.11$ kcal/mol.
On the other hand, the same calculations of {\bf 1b} ($R_\text{FeN}=2.048$ {\AA}), which is optimized for
the triplet state, result in quintet being more stable by 7.83 kcal/mol.
Note that although the {\bf 1b} model comes from the triplet optimized geometry, its Fe--N distance closely
reflects the quintet state geometry of FeP and its derivatives and the conclusions made on these systems will thus be slightly biased towards quintet (just as {\bf 1a} reflects their triplet
geometry and is biased towards triplet, see Table \ref{tab:geom}).
As can be seen from the comparison of the spin state ordering of {\bf 1} with FeP and FePc (see Table \ref{tab:dft}
and ESI), the former is not a sufficient model to describe the correlation of electrons in  Fe(3d,4d) orbitals
and pyrrolic $\pi$-electron system in Fe(II)-porphyrins.
In addition, the changes in triplet and quintet geometries of {\bf 1} are overestimated due to the increased
flexibility caused by removing the C$_\beta$H groups.
Despite this, our results highlight the crucial role of Fe--N distance in the spin-state ordering
and shed new light on the experimental data interpretation of Fe(II)-porphyrins.

% SUMMARY
In this communication, we presented a thorough study of Fe(II)-porphyrin model, which explored various effects influencing
the spin state ordering of FeP systems.
We included all previously discussed correlation effects\cite{LiManni2016,LiManni2018,LiManni2019,Zhou2019} -- non-dynamic
valence correlation via DMRG-CASSCF(32,34), and beyond-active-space and semi-core dynamic correlation
via DMRG-based DLPNO-TCCSD(T).
The use of the latter method allowed us to employ basis sets flexible enough to capture subtle changes in the
spin state ordering.
On top of that, we stress the crucial importance of geometrical parameters, the Fe--N distances
in particular, which is an aspect that has not been previously addressed and has a substantial impact
on the ground state character.
By exploring different geometries, we conclude that by using the model structure with Fe--N distances close
to the quintet optimized geometry of FeP and its derivatives, the ground state is found to be a quintet
(vertical $\Delta E^\mathrm{Q \rightarrow T} = 7.8$ kcal/mol), which is consistent with the previous
measurements on an isolated molecule of Fe(II)-phthalocyanine\cite{Nachtigallov2018}.

\section*{Conflicts of interest}
There are no conflicts to declare.

\section*{Acknowledgment}
This work has been supported by
the Czech Science Foundation (18-18940Y and 19-27454X),
Charles University (GAUK 376217),
Czech Ministry of Education (LTAUSA17033),
the Hungarian National Research, Development and Innovation Office (K120569)
and
the Hungarian Quantum Technology National Excellence Program (2017-1.2.1-NKP-2017-00001).
We acknowledge the IT4Innovations National Supercomputing Center.
%LV acknowledges support by the Czech Ministry of Education%, Youth and Sports
%( LTAUSA17033) and the IT4Innovations National Supercomputing Center
%Large Infrastructures for Research, Experimental Development
%and Innovations project ``IT4Innovations National Supercomputing Center -- LM2015070'').
\"{O}L acknowledges financial support from the Alexander von Humboldt foundation.
The development of the MOLMPS library was supported by the Center for Scalable and Predictive methods for Excitation and Correlated phenomena (SPEC), which is funded by the U.S. Department of Energy (DOE), Office of Science, Office of Basic Energy Sciences, the Division of Chemical Sciences, Geosciences, and Biosciences.
%%%END OF MAIN TEXT%%%

 %For footnotes in the main text of the article please number the footnotes to avoid duplicate symbols. e.g.  \footnote[num]{your text} the corresponding author \ast counts as footnote 1, ESI as footnote 2, e.g. if there is no ESI, please start at [num]=[2], if ESI is cited in the title please start at [num]=[3] etc. Please also cite the ESI within the main body of the text using \dag.

%The \balance command can be used to balance the columns on the final page if desired. It should be placed anywhere within the first column of the last page.

%\balance

%If notes are included in your references you can change the title from 'References' to 'Notes and references' using the following command:
%\renewcommand\refname{Notes and references}

%%%REFERENCES%%%
%\scriptsize{
\bibliography{references} %You need to replace "rsc" on this line with the name of your .bib file

\end{document}